\documentclass[aps,prd,twocolumn,superscriptaddress,floatfix]{revtex4}
\input epsf
\usepackage{graphics}

\begin{document}
\title{Filling the holes: Evolving excised binary black hole initial data with puncture techniques}
\date{\today}
\author{Zachariah B. Etienne}
\affiliation{Department of Physics, University of Illinois at
  Urbana-Champaign, Urbana, IL 61801}
\author{Joshua A. Faber}
\altaffiliation{National Science Foundation (NSF) Astronomy and
  Astrophysics Postdoctoral Fellow.}
\affiliation{Department of Physics, University of Illinois at
  Urbana-Champaign, Urbana, IL 61801}
\author{Yuk Tung Liu}
\affiliation{Department of Physics, University of Illinois at
  Urbana-Champaign, Urbana, IL 61801}
\author{Stuart~L.~Shapiro}
\altaffiliation{Also at Department of Astronomy and NCSA, University of
  Illinois at Urbana-Champaign, Urbana, IL 61801}
\affiliation{Department of Physics, University of Illinois at
  Urbana-Champaign, Urbana, IL 61801}
\author{Thomas W. Baumgarte}
\altaffiliation{Also at Department of Physics, University of Illinois at
  Urbana-Champaign, Urbana, IL 61801}
\affiliation{Department of Physics and Astronomy, Bowdoin College,
  Brunswick, ME 04011}

\begin{abstract}
We follow the inspiral and merger of equal-mass black holes (BHs) by the
moving puncture technique and demonstrate that both the exterior solution and the
asymptotic gravitational waveforms are unchanged when the initial
interior solution is replaced by constraint-violating ``junk'' initial
data.  We apply this result to evolve conformal thin-sandwich (CTS)
binary BH initial data by filling their excised interiors with
arbitrary, but smooth, initial data and evolving with standard
puncture gauge choices.
The waveforms generated for both puncture and filled-CTS initial data
are remarkably similar, and there are only minor differences between
irrotational and corotational CTS BH binaries.  Even the interior
solutions appear to evolve to the same constraint-satisfying solution
at late times, independent of the initial data.
\end{abstract}
\maketitle

\section{Introduction}

Numerical relativity
has made dramatic progress over the past two years in solving
the binary black hole (BBH) problem via the ``moving puncture'' technique
\cite{RIT1,God1,FP}.  To construct puncture 
initial data, the spatial metric is taken to be conformally flat and
the conformal factor is split into an algebraic singular
piece and a numerically determined regular contribution \cite{BB,Baum00}.
These initial data are then evolved
via the BSSN \cite{SN,BS} formulation of Einstein's equations, {\it
  without} special treatment of the initial (coordinate)
singularities at the punctures.  Coupled to the BSSN field evolution
equations are a set of gauge evolution equations, similar to those
discussed in \cite{BM,Alcu} and
successfully evolved by \cite{RIT1,God1}, that are critical for long-term stability.

One possible reservation
is that puncture initial data are thought to contain more inherent
eccentricity \cite{BIW} and, possibly, more initial spurious gravitational
radiation than those constructed using the conformal thin-sandwich 
(CTS) method.  CTS initial data are typically constructed
with equilibrium boundary conditions imposed on an excised surface 
defining the BH horizon
\cite{CP,Caudill,CC}, and it has 
been suggested that the lack of data in the BH interior may pose a
significant problem for evolving CTS initial data with puncture
techniques (see, e.g., \cite{Jena6th}).  
Thus, excision-based initial
data, for both orbiting binaries \cite{GodRITFP,AEIGeo} and head-on
collisions \cite{Sperhake1,Sperhake2}, have heretofore been evolved via
dynamical excision, rather than puncture techniques.
We will demonstrate here that there are no difficulties evolving
BH interiors artificially filled with smoothly extrapolated ``junk''
(i.e., constraint-violating) initial data via
puncture techniques: the puncture gauge conditions quickly drive
the evolution toward the usual late-time puncture interior solutions,
independent of the initial interior data.

This demonstration follows from several previous studies of BHs in puncture
gauges.
In \cite{JenaGeo1,JenaGeo2,BN}, it was shown that
the conformal factor $\psi$
evolves from a $1/r$ dependence at small $r$ to $1/\sqrt{r}$ at late
times, indicating that the spatial slices terminate on a limiting surface of finite
areal radius.
It was argued in \cite{Brown1} that this effect is really a consequence of
``excision via
under-resolution'', namely the fact that the puncture gauge conditions
make numerical grid points move away from the puncture very rapidly.  In a Kruskal
diagram, where the puncture is represented by the asymptotic end of
region III, all grid points quickly move from region III into region
II, leaving few or no grid points behind in the ``left" part of the
diagram (see Fig.~2 in \cite{Brown1}). 

In a recent paper \cite{FBEST}, we demonstrated that one can replace the solution inside
the horizon of a single {\it stationary} puncture BH by ``junk data''.
We find that this junk does not affect the BH exterior and, more
surprisingly, even in the black hole interior the BH settles down
to the same constraint-satisfying solution, independently of the
initial data.  This can again be motivated with the help of a Kruskal
diagram.  We impose junk on the $t=0$ slice in the black hole
interior, i.e.~in the ``left" part of the diagram.
As demonstrated by \cite{Brown1}, the puncture gauge conditions make all
grid points propagate toward the right at extremely rapid, ``superluminal'' speeds.
Assuming that constraint violations travel with much smaller speeds,
the grid points are likely to leave the junk's future domain
of dependence eventually, and enter a region that is affected by
constraint-satisfying initial data only. 

Here, we extend our work to the BBH case.
We first demonstrate numerically that replacing interior puncture
initial data by ``junk'' data does not alter the exterior evolution.
Given this finding, we then
show that the excised CTS initial data generated by Cook
and Pfeiffer (\cite{CP}; hereafter CP) can be evolved using puncture
gauges without loss of accuracy by filling in the interior
with arbitrary, but smooth, initial data. 
Rather than producing highly accurate waveforms,
the goal of this paper is to
establish a point of principle about initial data for dynamical 
evolutions.

\section{Initial Data}\label{sec:ID}
Throughout this paper we cast the spacetime metric $g_{ab}$ into its
3+1 ADM form 
\begin{equation}
ds^2=-(\alpha^2-\beta_i\beta^i)dt^2+2\beta_i dt dx^i +
\psi^4\tilde{\gamma}_{ij}dx^idx^j,
\end{equation}
with lapse function $\alpha$, shift vector $\beta^i$, and spatial metric 
$\gamma_{ij} = \psi^{4}\tilde{\gamma}_{ij}$, where
$\tilde{\gamma}_{ij}$ is the conformally related metric and the 
conformal factor $\psi$ is chosen so that
$\det(\tilde{\gamma})=1$ in Cartesian coordinates. 
Both puncture and CP spatial initial data are 
conformally flat at $t=0$, i.e., $\tilde{\gamma}_{ij}=\eta_{ij}$.

In generating BBH puncture initial data~\cite{BB,Baum00}, the momentum
constraint reduces to an algebraic expression and the Hamiltonian
constraint to a nonlinear elliptic equation governing the nonsingular
part of $\psi$, which we solve via the {\tt Lorene} multidomain
spectral methods libraries~\cite{Lorene}.  These constraint equations
depend on the initial linear momenta, coordinate separation, spins,
and puncture masses.  We obtain these initial
parameters from the quasiequilibrium, circular sequence of~\cite{TB},
first treating the case of nonspinning, equal-mass BHs.

To better understand the role played by the BH interiors in 
puncture evolutions, we replace the puncture interiors (``P'') with ``junk'' initial
data.  Specifically, we replace
the singular term of $\psi$ with some other smooth
function inside a radius $r_J$ and leave the nonsingular part
untouched, as we did in \cite{FBEST} for stationary BHs.

The singular part of $\psi$ is 
$\psi_s = \sum_i(M_i/2r_i)$,
where $M_i$ is the puncture mass of BH $i$, and
$r_i$ is the distance from the origin to BH $i$.  We construct {\it smooth
  junk} (``${\rm P}_{\rm SJ}$'') by replacing the term $M_i/2r_i$ by an even,
fourth-order polynomial inside $r_J/M_i = 0.4$ (in these isotropic
coordinates, the apparent horizon radius is located at $r_{\rm AH}\approx 0.5M_i$
initially), chosen
such that the resulting conformal factor is continuous and twice
differentiable everywhere.  {\it Flat junk} (``${\rm P}_{\rm FJ}$'') is defined by replacing
$M_i/2r_i$ with a constant value $2.0$ within $r_J/M_i = 0.25$.  This
second choice produces a conformal 
factor that is continuous, but not differentiable everywhere.  Clearly, the
the Hamiltonian constraint is {\em not} satisfied within the junk
regions of our grid initially.  

The corotating CP initial data (``${\rm CP}_{\rm C}$'') are available
from \cite{bhorg}, whereas the irrotational
CP data (``${\rm CP}_{\rm I}$'') were graciously supplied by Harald Pfeiffer.  
Unlike puncture initial data, which are defined everywhere in
space, CP initial data are produced by imposing equilibrium boundary
conditions on excision surfaces -- chosen to be coordinate spheres --  
that define the BH's apparent horizons.
This procedure produces valid data only in the BH exterior.  To aide in some
numerical simulations, the CP initial data were provided with data extrapolated from
the BH exterior to a region outside 
$0.75r_{\rm AH}$.  To
fill the remaining excised region inside $0.75 r_{\rm AH}$, we
extrapolate all field values radially from $0.8r_{\rm AH}$ to the BH
center, matching smoothly to fourth-order polynomials with a fixed, finite
value at $r=0$, so
that the resulting field variables are everywhere continuous and twice
differentiable, similar to the ${\rm P}_{\rm SJ}$ puncture case. 
Evolving with the CTS gauge variables $\alpha$ and $\beta^i$ 
results in eccentric BH coordinate trajectories.  
We can remove this gauge-dependent eccentricity considerably by choosing instead the
same initial 
lapse and shift as in the puncture cases, $\alpha = \psi^{-2}$ and
$\beta^i=0$, evolving with Eqs.~(\ref{eq:gauge1}) --
(\ref{eq:gauge2}) below.

\begin{table}
\caption{Summary of parameters for our simulations.  For initial
    data, ``CP'' refers to CTS Cook-Pfeiffer data, smoothly extrapolated
    inside the excised regions. $M$ is the total ADM mass of the binary
    system, $r_0$ and $J_0$ the initial separation and total
    angular momentum, and $M_{irr}$ the irreducible mass of each BH.
    $\Delta E$ and $\Delta J$ are the total radiated energy and angular
    momentum during the course of the evolution in the dominant $l=2$,
    $m=\pm 2$ modes, evaluated at extraction radii of $17.8M$. }
\begin{tabular}{ll|llll|ll}
\hline
Name & Init. Data & $M\Omega$ & $r_0/M$ & $J_0/M^2$ &
$M_{irr}/M$ & $\Delta E/M$ & $\Delta J/J_0$ \\
\hline\hline
P & No Junk & 0.0824 & 2.185 & 0.807 & 0.476 & 0.026 & 0.16 \\
${\rm P}_{\rm SJ}$ & Sm. Junk & 0.0824 & 2.185 & 0.807 & 0.476 & 0.027 & 0.17  \\
${\rm P}_{\rm FJ}$ & Flat Junk & 0.0824 & 2.185 & 0.807 & 0.476 & 0.026 & 0.16  \\
\hline
${\rm CP}_{\rm I}$ & Irrot. CP & 0.0824 & 2.237 & 0.823 & 0.447 & 0.029 & 0.19  \\
${\rm CP}_{\rm C}$ & Corot. CP & 0.0817 & 2.222 & 0.877 & 0.444 & 0.028 & 0.19   \\
\hline
\end{tabular}
\label{table:runs}
\end{table}

In Table~\ref{table:runs}, we summarize the parameters for each of our
numerical simulations.  Both irrotational and corotating CP data are evolved
from an initial separation $r_0\approx 2.2M$, which is beyond the
quasiequilibrium ISCO radius, and comparable to the initial
configurations evolved in \cite{RIT1,God1}.  Here and below
$M$ is the binary's total ADM mass.  For comparison, we
construct puncture initial data with the same orbital frequency as the
irrotational CP data ($M\Omega=0.08241$), using the equal-mass
irrotational fitting formulae of \cite{TB} to determine our initial
parameters.  All runs were performed on $420^2\times 210$ spatial
grids, assuming equatorial symmetry.  We employ
multiple-transition fisheye coordinates \cite{RIT1} to expand the
physical extent of our grid, as we did in \cite{FBEST}.  Our grid
includes three 1:4 fisheye transitions at $r/M$ = 3.89, 5.22, and 6.55,
all with transition width $s/M=0.417$.  Resolution in
the innermost region is set to $M/24$, so the outer boundary falls at
a physical distance of $\approx 170M$.

\section{Dynamical Calculations}\label{sec:runs}
Our evolutions are
performed using the same Cactus-based \cite{Cactus}
GRMHD code described in \cite{FBEST}, but
with the MHD matter and E\&M field sectors disabled.  Our originally
second-order Iterative Crank-Nicholson BSSN-based code has been
upgraded to fourth-order accurate spatial 
differencing with upwind differencing for advective terms, but we still
use second-order differencing for the time evolution.

We use standard puncture gauge evolution equations, specifically
a $1+\log$ lapse and a ``non-shifting shift'':
\begin{eqnarray}
\partial_t \alpha&=&\beta^j\partial_j\alpha -2\alpha K\label{eq:gauge1}\\
\partial_t \beta^i&=&(3/4)B^i;~\partial_t B^i=\partial_t \tilde{\Gamma}^i-\eta B^i\label{eq:gauge2}
\end{eqnarray}
where we set $\eta=0.5/M$ in all of our runs.

\begin{figure}[ht!]
 \centering \leavevmode \epsfxsize=2.45in \epsfbox{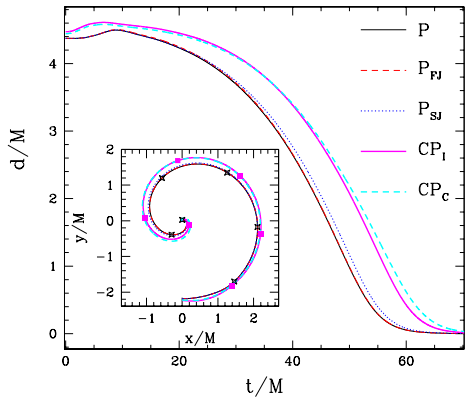}
 \caption{Binary separation versus time and paths traced in coordinate
 space (inset) for runs P
 (thin solid), ${\rm P}_{\rm FJ}$ (thin dashed), ${\rm P}_{\rm SJ}$ (thin dotted), 
 ${\rm CP}_{\rm I}$ (thick solid), and ${\rm CP}_{\rm C}$ (thick dashed).  In the inset, we show
 the motion of one of the BHs in the equatorial plane, with points representing
 positions every $10M$ for runs P (open squares) and ${\rm CP}_{\rm
 I}$ (closed squares).  While $\pi$-symmetry is not enforced, it holds
 true numerically during our evolutions.}
\label{fig:xbh}
\end{figure}

In Fig.~\ref{fig:xbh}, we show the coordinate trajectories of one of the
punctures, which obey the relation
$dx^i/dt=-\beta^i(x^i)$\cite{RIT1}.  For runs with non-singular junk
data in the interior, we locate the initial BH position at $r=0$,
coincident with the standard puncture case.
While the puncture and CP runs
trace out slightly different tracks, this comparison is 
gauge-dependent. 
To evaluate our runs in an asymptotically gauge-invariant manner, we compute the
gravitational waveforms through the Weyl scalar $\psi_4$ via
the Cactus
{\tt PsiKadelia} thorn \cite{PsiK}, which we have modified to make the
computation of spatial derivatives accurate to fourth order.  We have
tested our modified PsiKadelia thorn with a full suite of analytic linearized
wave tests.  The total energy and angular 
momentum lost to gravitational radiation are computed by integrating Eqs.~(5.1) and
(5.2) in \cite{BCLT}, keeping only the 
$s=-2$ spin-weighted, $l=2$, $m=\pm 2$ contributions from $\psi_4$ to reduce numerical noise.
We have performed simulations at various resolutions using the
same puncture initial data as in \cite{RIT1}, finding that our waveform
agrees with Fig.~3 of the preprint version of \cite{RIT1} to high accuracy, and converges to
second order, as we show in Fig.~\ref{fig:gw2} for the ${\rm CP}_{\rm
  I}$ case.

\begin{figure}[ht!]
 \centering \leavevmode \epsfxsize=2.45in \epsfbox{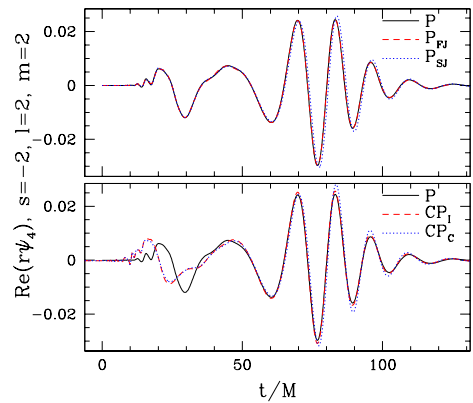}
 \caption{Real part of the $s=-2$ spin-weighted, $l=2$, $m=2$
 component of the Weyl scalar $\psi_4$, evaluated on a sphere of
 radius $r=17.8M$.  In the top panel we show results for puncture runs P
 (solid), ${\rm P}_{\rm FJ}$ (dashed), and ${\rm P}_{\rm SJ}$ (dotted), which differ in the form of
 the initial junk we impose within each BH horizon.  
 In the bottom panel, we compare run P with our two runs started from
 CP initial data, both irrotational (dashed) and corotating
 (dotted).  We see extremely close agreement between puncture
 and irrotational CP results, and only minor differences between
 these and the corotating CP results.}
\label{fig:gw}
\end{figure}

\begin{figure}[ht!]
 \centering \leavevmode \epsfxsize=2.45in \epsfbox{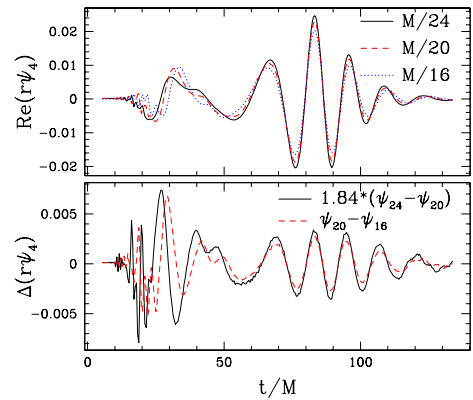}
 \caption{Waveforms for runs using the initial data of run ${\rm CP}_{\rm
  I}$ (top panel) evaluated at numerical resolutions of $M/24$ (solid), $M/20$
  (dashed), and $M/16$ (dotted).  Results are time and phase shifted
  so that the moment and phase of the maximum wave amplitudes
  coincide.  In the bottom panel, we show the differences between the
  runs, properly rescaled to highlight second-order convergence.}
\label{fig:gw2}
\end{figure}

In~\cite{FBEST}, we demonstrated that even if single stationary punctures are
constructed with different junk initial interior data, the ADM mass
versus time and conformal factor profiles at late times 
agree to high accuracy.  Here, we demonstrate that the
gravitational waveforms generated by puncture BBHs are also largely independent
of interior junk.  In Fig.~\ref{fig:gw},
we show the real 
$l=2$, $m=2$-mode component of $\psi_4$, evaluated on a sphere
of physical (as opposed to fisheye) radius $r=17.8M$, comparable to
the extraction radius used in \cite{RIT1,God1}.  In the top panel, 
we show a comparison
between simulations P, ${\rm P}_{\rm FJ}$, and ${\rm P}_{\rm SJ}$.  We find
that the first two cases are almost identical, even though the data ${\rm
  P}_{\rm FJ}$  are not differentiable initially 
in the BH interior.  There is a slight phase difference for run ${\rm
  P}_{\rm SJ}$, 
since the differencing stencil in the inner edge of the exterior
overlaps junk data in the interior that do not initially satisfy the
constraints. In the bottom panel of the figure, we compare results from simulation P
with those from CP initial data (the CP curves are time
and phase-shifted equally so that the time and phase at maximum
amplitude match between runs P and ${\rm CP}_{\rm I}$).  We note that
after an early 
phase of transient gravitational radiation propagation, the waveforms
from the merger and ringdown of runs P and ${\rm CP}_{\rm I}$ match very
closely, varying in amplitude by no more than $4\%$ with nearly
exact phase agreement.  These relative differences are well within the relative
differences in the initial data's masses and angular momenta, as listed in
Table~\ref{table:runs}.

\begin{figure}[ht!]
 \centering \leavevmode \epsfxsize=2.45in \epsfbox{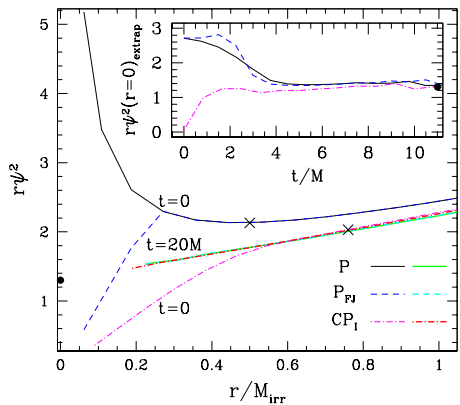}
 \caption{Radial profiles of the quantity $r\psi^2$ as a function
 of coordinate radius, measured out from
 the BH center for runs P and ${\rm CP}_{\rm I}$.  The evolution 
inevitably drives $\psi$ toward the power-law index $\psi\propto
 r^{-1/2}$ derived in \protect\cite{JenaGeo1}.  For the curves showing
 profiles at $t=20M$, we eliminate the two innermost points, whose
 values are inaccurate due to differencing across the puncture.
Crosses denote the initial apparent horizons for each run.
In the inset, we plot the value of $r\psi^2$ linearly extrapolated
 from the third and fourth innermost data points as a function of time for the
 same runs. The filled circles show the asymptotic value $1.3M$
 determined analytically by \protect\cite{JenaGeo1}. } 
\label{fig:psir}
\end{figure}

The similarity between puncture and CP merger waveforms may not be
particularly surprising, since the gauge evolution equations
inevitably drive a BH 
toward the universal form described in \cite{JenaGeo1}.  To
demonstrate this effect, in Fig.~\ref{fig:psir} we compare
profiles of $r\psi^2$
for runs P and ${\rm CP}_{\rm I}$ at $t=0$ and $20M$, before coalescence.  
In both cases, the gauge conditions we choose
drive the solution toward
$\psi\propto r^{-1/2}$, regardless of whether the initial data are
singular or expressly finite. 
For an isolated BH, it was shown in \cite{JenaGeo1} that the puncture $r=0$ evolves
to a limiting slice of finite Schwarzschild radius,
$r_s=\psi^2r\approx 1.3M$.  We observe the same asymptotic value for
this function in
all of our binary evolutions prior to coalescence.

\begin{figure}[ht!]
 \centering \leavevmode \epsfxsize=2.45in \epsfbox{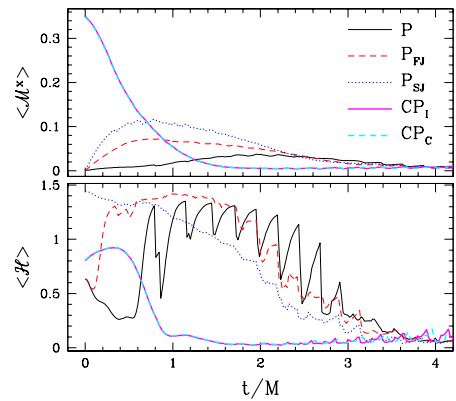}
 \caption{Violation of the Hamiltonian (${\mathcal
 H}=0$) and momentum (${\mathcal M}^i=0$) constraints.  We show the
 normalized x-component of the momentum constraint violation 
$<{\mathcal M}^x>\equiv \sqrt{\sum({\mathcal M}^x)^2}/N_{MC}$ (top
 panel), where $N_{MC}$ is the $L_2$-norm of ${\mathcal M}^x$ given by
 Eq.~60 of \protect\cite{DMSB}, and a similar quantity for the
 Hamiltonian constraint (bottom panel).  These are summed over
 spherical shells about each puncture
 $0.2<r/M<0.3$, which extend into the BH interiors.  The
 constraint-violating initial data are driven to constraint-satisfying
 solutions at late times (several $M$).}
\label{fig:cons}
\end{figure}

In Fig.~\ref{fig:cons}, we compare the magnitude of the normalized
constraint violations \cite{DMSB}
for different initial data sets, summing over spherical shells about each
puncture spanning
$0.2<r/M<0.3$.  These shells allow us to study constraint violations
propagating outward in the immediate vicinity of the BHs, the exterior region most
sensitive to the use of interior junk.
We find a brief but 
transient increase in the magnitude of the violations at early times in
all the puncture simulations, whereas the CP simulations begin with larger
violations but are driven toward constraint-satisfying configurations quite rapidly.

\section{Discussion}\label{sec:discussion}

For solutions of Einstein's equations, the BH exterior must, by
definition, be independent 
of the BH interior.  As long as constraint violations propagate
``causally", this must also be
true if the BH interior violates the constraints  (see the
subsequent discussion in \cite{turducken}).  Previous analyses
\cite{JenaGeo1,JenaGeo2,BN,Brown1} also suggest that, with the
``moving puncture" gauges, any initial BH configuration would be
driven to the asymptotic puncture solution found in \cite{JenaGeo1}.
We demonstrate numerically that this is indeed the case and show that
one can modify the data within the BH interior and produce valid 
waveforms.  The success of the puncture method does not depend on
the form of the initial data, or even on the existence of valid
initial data, in the BH interior. 

More specifically, we find that evolutions of puncture initial data
are insensitive to interior ``junk'' so long as the finite differencing
stencils of exterior points do not overlap regions where constraints
are violated.  One may freely modify the interior
fields to eliminate initially asymptotic singular behavior, or even invent initial
data in the interior if they do not exist, so long as
the solution matches smoothly to the exterior so that all
fields are twice differentiable and satisfy the constraints in regions
covered by the differencing stencils of exterior points.

As a consequence, 
excised initial data, in particular the CTS sets constructed in
\cite{CP,Caudill,CC}, may be evolved using puncture
techniques.  These data produce a different initial burst of spurious
gravitational radiation, reflecting small differences in the ``imperfection'' of the initial data, but evolve similarly to punctures thereafter.
This is true for both irrotational and corotating CTS data.  The
differences are small since the
latter involve only marginally spinning BHs ($S/M_{\rm irr}^2=0.15$).
In fact, all our waveforms are very similar, and any relative differences in the 
wave signals are well within the relative differences of the physical parameters 
describing the initial data.

These results are also highly relevant for black hole-neutron star
(BHNS) binary evolutions.  To date, most published relativistic
quasiequilibrium 
BHNS sequences are calculated using the CTS method and excision \cite{TBFS1,TBFS2,Grand}
(the exception are the puncture data constructed in \cite{STU1,STU2}).  As demonstrated in this 
paper, these data can nevertheless be evolved with the moving puncture
technique by appropriately filling the excised region.

Finally, we note that the interior metric seems to evolve to the same
asymptotic form found by \cite{JenaGeo1}, independently of the initial
interior data.  Consequently, the late-time interior solutions on the
numerical grid satisfy the constraints with increasing accuracy, even
if the initial data consist of constraint-violating ``junk'' (see Fig.~\ref{fig:cons}).  This
feature presumably helps stabilize the integrations and explains the
robustness of the puncture approach.

We are very grateful to H.~Pfeiffer for providing
us with irrotational CTS Cook-Pfeiffer initial data.
We also thank G.~Cook and Y.~Zlochower
for helpful conversations.  JAF is
supported by an NSF Astronomy and Astrophysics Postdoctoral Fellowship
under award AST-0401533.  This work was supported in part by NSF
grants PHY-0205155 and PHY-0345151 and NASA Grants NNG04GK54G and
NNX07AG96G to the University of Illinois, and NSF Grant PHY-0456917 to
Bowdoin College.  All simulations were performed on the NCSA {\tt abe}
cluster.

\bibliography{paper4bib}

\begin{thebibliography}{32}
\expandafter\ifx\csname natexlab\endcsname\relax\def\natexlab#1{#1}\fi
\expandafter\ifx\csname bibnamefont\endcsname\relax
  \def\bibnamefont#1{#1}\fi
\expandafter\ifx\csname bibfnamefont\endcsname\relax
  \def\bibfnamefont#1{#1}\fi
\expandafter\ifx\csname citenamefont\endcsname\relax
  \def\citenamefont#1{#1}\fi
\expandafter\ifx\csname url\endcsname\relax
  \def\url#1{\texttt{#1}}\fi
\expandafter\ifx\csname urlprefix\endcsname\relax\def\urlprefix{URL }\fi
\providecommand{\bibinfo}[2]{#2}
\providecommand{\eprint}[2][]{\url{#2}}

\bibitem[{\citenamefont{{Campanelli} et~al.}(2006)\citenamefont{{Campanelli},
  {Lousto}, {Marronetti}, and {Zlochower}}}]{RIT1}
\bibinfo{author}{\bibfnamefont{M.}~\bibnamefont{{Campanelli}}},
  \bibinfo{author}{\bibfnamefont{C.~O.} \bibnamefont{{Lousto}}},
  \bibinfo{author}{\bibfnamefont{P.}~\bibnamefont{{Marronetti}}},
  \bibnamefont{and}
  \bibinfo{author}{\bibfnamefont{Y.}~\bibnamefont{{Zlochower}}},
  \bibinfo{journal}{\prl} \textbf{\bibinfo{volume}{96}},
  \bibinfo{pages}{111101} (\bibinfo{year}{2006}).

\bibitem[{\citenamefont{{Baker} et~al.}(2006)\citenamefont{{Baker},
  {Centrella}, {Choi}, {Koppitz}, and {van Meter}}}]{God1}
\bibinfo{author}{\bibfnamefont{J.~G.} \bibnamefont{{Baker}}},
  \bibinfo{author}{\bibfnamefont{J.}~\bibnamefont{{Centrella}}},
  \bibinfo{author}{\bibfnamefont{D.-I.} \bibnamefont{{Choi}}},
  \bibinfo{author}{\bibfnamefont{M.}~\bibnamefont{{Koppitz}}},
  \bibnamefont{and} \bibinfo{author}{\bibfnamefont{J.}~\bibnamefont{{van
  Meter}}}, \bibinfo{journal}{\prl} \textbf{\bibinfo{volume}{96}},
  \bibinfo{pages}{111102} (\bibinfo{year}{2006}).

\bibitem[{\citenamefont{Pretorius}(2005)}]{FP}
\bibinfo{note}{For an alternate method, see}
\bibinfo{author}{\bibfnamefont{F.}~\bibnamefont{{Pretorius}}},
  \bibinfo{journal}{\prl} \textbf{\bibinfo{volume}{95}},
  \bibinfo{pages}{121101} (\bibinfo{year}{2005}).

\bibitem[{\citenamefont{{Brandt} and {Br{\"u}gmann}}(1997)}]{BB}
\bibinfo{author}{\bibfnamefont{S.}~\bibnamefont{{Brandt}}} \bibnamefont{and}
  \bibinfo{author}{\bibfnamefont{B.}~\bibnamefont{{Br{\"u}gmann}}},
  \bibinfo{journal}{\prl} \textbf{\bibinfo{volume}{78}}, \bibinfo{pages}{3606}
  (\bibinfo{year}{1997}).

\bibitem[{\citenamefont{{Baumgarte}}(2000)}]{Baum00}
\bibinfo{author}{\bibfnamefont{T.~W.} \bibnamefont{{Baumgarte}}},
  \bibinfo{journal}{\prd} \textbf{\bibinfo{volume}{62}},
  \bibinfo{pages}{024018} (\bibinfo{year}{2000}).

\bibitem[{\citenamefont{{Shibata} and {Nakamura}}(1995)}]{SN}
\bibinfo{author}{\bibfnamefont{M.}~\bibnamefont{{Shibata}}} \bibnamefont{and}
  \bibinfo{author}{\bibfnamefont{T.}~\bibnamefont{{Nakamura}}},
  \bibinfo{journal}{\prd} \textbf{\bibinfo{volume}{52}}, \bibinfo{pages}{5428}
  (\bibinfo{year}{1995}).

\bibitem[{\citenamefont{{Baumgarte} and {Shapiro}}(1999)}]{BS}
\bibinfo{author}{\bibfnamefont{T.~W.} \bibnamefont{{Baumgarte}}}
  \bibnamefont{and} \bibinfo{author}{\bibfnamefont{S.~L.}
  \bibnamefont{{Shapiro}}}, \bibinfo{journal}{\prd}
  \textbf{\bibinfo{volume}{59}}, \bibinfo{pages}{024007}
  (\bibinfo{year}{1999}).

\bibitem[{\citenamefont{{Bona} et~al.}(1995)\citenamefont{{Bona}, {Mass{\'o}},
  {Seidel}, and {Stela}}}]{BM}
\bibinfo{author}{\bibfnamefont{C.}~\bibnamefont{{Bona}}},
  \bibinfo{author}{\bibfnamefont{J.}~\bibnamefont{{Mass{\'o}}}},
  \bibinfo{author}{\bibfnamefont{E.}~\bibnamefont{{Seidel}}}, \bibnamefont{and}
  \bibinfo{author}{\bibfnamefont{J.}~\bibnamefont{{Stela}}},
  \bibinfo{journal}{Physical Review Letters} \textbf{\bibinfo{volume}{75}},
  \bibinfo{pages}{600} (\bibinfo{year}{1995}), \eprint{arXiv:gr-qc/9412071}.

\bibitem[{\citenamefont{{Alcubierre} et~al.}(2003)\citenamefont{{Alcubierre},
  {Br{\"u}gmann}, {Diener}, {Koppitz}, {Pollney}, {Seidel}, and
  {Takahashi}}}]{Alcu}
\bibinfo{author}{\bibfnamefont{M.}~\bibnamefont{{Alcubierre}}},
  \bibinfo{author}{\bibfnamefont{B.}~\bibnamefont{{Br{\"u}gmann}}},
  \bibinfo{author}{\bibfnamefont{P.}~\bibnamefont{{Diener}}},
  \bibinfo{author}{\bibfnamefont{M.}~\bibnamefont{{Koppitz}}},
  \bibinfo{author}{\bibfnamefont{D.}~\bibnamefont{{Pollney}}},
  \bibinfo{author}{\bibfnamefont{E.}~\bibnamefont{{Seidel}}}, \bibnamefont{and}
  \bibinfo{author}{\bibfnamefont{R.}~\bibnamefont{{Takahashi}}},
  \bibinfo{journal}{\prd} \textbf{\bibinfo{volume}{67}},
  \bibinfo{pages}{084023} (\bibinfo{year}{2003}), \eprint{arXiv:gr-qc/0206072}.

\bibitem[{\citenamefont{{Berti} et~al.}(2006)\citenamefont{{Berti}, {Iyer}, and
  {Will}}}]{BIW}
\bibinfo{author}{\bibfnamefont{E.}~\bibnamefont{{Berti}}},
  \bibinfo{author}{\bibfnamefont{S.}~\bibnamefont{{Iyer}}}, \bibnamefont{and}
  \bibinfo{author}{\bibfnamefont{C.~M.} \bibnamefont{{Will}}},
  \bibinfo{journal}{\prd} \textbf{\bibinfo{volume}{74}},
  \bibinfo{pages}{061503} (\bibinfo{year}{2006}).

\bibitem[{\citenamefont{{Cook} and {Pfeiffer}}(2004)}]{CP}
\bibinfo{author}{\bibfnamefont{G.~B.} \bibnamefont{{Cook}}} \bibnamefont{and}
  \bibinfo{author}{\bibfnamefont{H.~P.} \bibnamefont{{Pfeiffer}}},
  \bibinfo{journal}{\prd} \textbf{\bibinfo{volume}{70}},
  \bibinfo{pages}{104016} (\bibinfo{year}{2004}).

\bibitem[{\citenamefont{{Caudill} et~al.}(2006)\citenamefont{{Caudill}, {Cook},
  {Grigsby}, and {Pfeiffer}}}]{Caudill}
\bibinfo{author}{\bibfnamefont{M.}~\bibnamefont{{Caudill}}},
  \bibinfo{author}{\bibfnamefont{G.~B.} \bibnamefont{{Cook}}},
  \bibinfo{author}{\bibfnamefont{J.~D.} \bibnamefont{{Grigsby}}},
  \bibnamefont{and} \bibinfo{author}{\bibfnamefont{H.~P.}
  \bibnamefont{{Pfeiffer}}}, \bibinfo{journal}{\prd}
  \textbf{\bibinfo{volume}{74}}, \bibinfo{pages}{064011}
  (\bibinfo{year}{2006}).

\bibitem[{\citenamefont{{Pfeiffer} et~al.}(2007)\citenamefont{{Pfeiffer},
  {Brown}, {Kidder}, {Lindblom}, {Lovelace}, and {Scheel}}}]{CC}
\bibinfo{author}{\bibfnamefont{H.~P.} \bibnamefont{{Pfeiffer}}},
  \bibinfo{author}{\bibfnamefont{D.~A.} \bibnamefont{{Brown}}},
  \bibinfo{author}{\bibfnamefont{L.~E.} \bibnamefont{{Kidder}}},
  \bibinfo{author}{\bibfnamefont{L.}~\bibnamefont{{Lindblom}}},
  \bibinfo{author}{\bibfnamefont{G.}~\bibnamefont{{Lovelace}}},
  \bibnamefont{and} \bibinfo{author}{\bibfnamefont{M.~A.}
  \bibnamefont{{Scheel}}}, \bibinfo{journal}{Class. Quant. Grav.}
  \textbf{\bibinfo{volume}{24}}, \bibinfo{pages}{59} (\bibinfo{year}{2007}).

\bibitem[{\citenamefont{{Husa} et~al.}(2007)\citenamefont{{Husa}, {Gonzalez},
  {Hannam}, {Bruegmann}, and {Sperhake}}}]{Jena6th}
\bibinfo{author}{\bibfnamefont{S.}~\bibnamefont{{Husa}}},
  \bibinfo{author}{\bibfnamefont{J.~A.} \bibnamefont{{Gonzalez}}},
  \bibinfo{author}{\bibfnamefont{M.}~\bibnamefont{{Hannam}}},
  \bibinfo{author}{\bibfnamefont{B.}~\bibnamefont{{Bruegmann}}},
  \bibnamefont{and}
  \bibinfo{author}{\bibfnamefont{U.}~\bibnamefont{{Sperhake}}},
  \bibinfo{journal}{ArXiv e-prints}  (\bibinfo{year}{2007}),
  \eprint{0706.0740}.

\bibitem[{\citenamefont{{Baker} et~al.}(2007)\citenamefont{{Baker},
  {Campanelli}, {Pretorius}, and {Zlochower}}}]{GodRITFP}
\bibinfo{author}{\bibfnamefont{J.~G.} \bibnamefont{{Baker}}},
  \bibinfo{author}{\bibfnamefont{M.}~\bibnamefont{{Campanelli}}},
  \bibinfo{author}{\bibfnamefont{F.}~\bibnamefont{{Pretorius}}},
  \bibnamefont{and}
  \bibinfo{author}{\bibfnamefont{Y.}~\bibnamefont{{Zlochower}}},
  \bibinfo{journal}{Class.~Quant.~Grav.} \textbf{\bibinfo{volume}{24}},
  \bibinfo{pages}{25} (\bibinfo{year}{2007}).

\bibitem[{\citenamefont{{Thornburg} et~al.}(2007)\citenamefont{{Thornburg},
  {Diener}, {Pollney}, {Rezzolla}, {Schnetter}, {Seidel}, and
  {Takahashi}}}]{AEIGeo}
\bibinfo{author}{\bibfnamefont{J.}~\bibnamefont{{Thornburg}}},
  \bibinfo{author}{\bibfnamefont{P.}~\bibnamefont{{Diener}}},
  \bibinfo{author}{\bibfnamefont{D.}~\bibnamefont{{Pollney}}},
  \bibinfo{author}{\bibfnamefont{L.}~\bibnamefont{{Rezzolla}}},
  \bibinfo{author}{\bibfnamefont{E.}~\bibnamefont{{Schnetter}}},
  \bibinfo{author}{\bibfnamefont{E.}~\bibnamefont{{Seidel}}}, \bibnamefont{and}
  \bibinfo{author}{\bibfnamefont{R.}~\bibnamefont{{Takahashi}}},
  \bibinfo{journal}{ArXiv e-prints}  (\bibinfo{year}{2007}),
  \eprint{gr-qc/0701038}.

\bibitem[{\citenamefont{{Sperhake}}(2006)}]{Sperhake1}
\bibinfo{author}{\bibfnamefont{U.}~\bibnamefont{{Sperhake}}},
  \bibinfo{journal}{ArXiv e-prints}  (\bibinfo{year}{2006}),
  \eprint{gr-qc/0606079}.

\bibitem[{\citenamefont{{Sperhake} et~al.}(2007)\citenamefont{{Sperhake},
  {Bruegmann}, {Gonzalez}, {Hannam}, and {Husa}}}]{Sperhake2}
\bibinfo{author}{\bibfnamefont{U.}~\bibnamefont{{Sperhake}}},
  \bibinfo{author}{\bibfnamefont{B.}~\bibnamefont{{Bruegmann}}},
  \bibinfo{author}{\bibfnamefont{J.}~\bibnamefont{{Gonzalez}}},
  \bibinfo{author}{\bibfnamefont{M.}~\bibnamefont{{Hannam}}}, \bibnamefont{and}
  \bibinfo{author}{\bibfnamefont{S.}~\bibnamefont{{Husa}}},
  \bibinfo{journal}{ArXiv e-prints}  (\bibinfo{year}{2007}),
  \eprint{0705.2035}.

\bibitem[{\citenamefont{{Hannam} et~al.}(2006)\citenamefont{{Hannam}, {Husa},
  {Pollney}, {Br\"ugmann}, and {O'Murchadha}}}]{JenaGeo1}
\bibinfo{author}{\bibfnamefont{M.}~\bibnamefont{{Hannam}}},
  \bibinfo{author}{\bibfnamefont{S.}~\bibnamefont{{Husa}}},
  \bibinfo{author}{\bibfnamefont{D.}~\bibnamefont{{Pollney}}},
  \bibinfo{author}{\bibfnamefont{B.}~\bibnamefont{{Br\"ugmann}}},
  \bibnamefont{and}
  \bibinfo{author}{\bibfnamefont{N.}~\bibnamefont{{O'Murchadha}}},
  \bibinfo{journal}{ArXiv e-prints}  (\bibinfo{year}{2006}),
  \eprint{gr-qc/0606099}.

\bibitem[{\citenamefont{{Hannam} et~al.}(2007)\citenamefont{{Hannam}, {Husa},
  {Br{\"u}gmann}, {Gonz{\'a}lez}, {Sperhake}, and {Murchadha}}}]{JenaGeo2}
\bibinfo{author}{\bibfnamefont{M.}~\bibnamefont{{Hannam}}},
  \bibinfo{author}{\bibfnamefont{S.}~\bibnamefont{{Husa}}},
  \bibinfo{author}{\bibfnamefont{B.}~\bibnamefont{{Br{\"u}gmann}}},
  \bibinfo{author}{\bibfnamefont{J.~A.} \bibnamefont{{Gonz{\'a}lez}}},
  \bibinfo{author}{\bibfnamefont{U.}~\bibnamefont{{Sperhake}}},
  \bibnamefont{and} \bibinfo{author}{\bibfnamefont{N.~{\'O}.}
  \bibnamefont{{Murchadha}}}, \bibinfo{journal}{J. Phys. Conf. Series}
  \textbf{\bibinfo{volume}{66}}, \bibinfo{pages}{2047} (\bibinfo{year}{2007}).

\bibitem[{\citenamefont{{Baumgarte} and {Naculich}}(2007)}]{BN}
\bibinfo{author}{\bibfnamefont{T.~W.} \bibnamefont{{Baumgarte}}}
  \bibnamefont{and} \bibinfo{author}{\bibfnamefont{S.~G.}
  \bibnamefont{{Naculich}}}, \bibinfo{journal}{\prd}
  \textbf{\bibinfo{volume}{75}}, \bibinfo{pages}{067502}
  (\bibinfo{year}{2007}).

\bibitem[{\citenamefont{{Brown}}(2007)}]{Brown1}
\bibinfo{author}{\bibfnamefont{J.~D.} \bibnamefont{{Brown}}},
  \bibinfo{journal}{ArXiv e-prints}  (\bibinfo{year}{2007}),
  \eprint{0705.1359}.

\bibitem[{\citenamefont{{Faber} et~al.}(2007)\citenamefont{{Faber},
  {Baumgarte}, {Etienne}, {Shapiro}, and {Taniguchi}}}]{FBEST}
\bibinfo{author}{\bibfnamefont{J.~A.} \bibnamefont{{Faber}}},
  \bibinfo{author}{\bibfnamefont{T.~W.} \bibnamefont{{Baumgarte}}},
  \bibinfo{author}{\bibfnamefont{Z.~B.} \bibnamefont{{Etienne}}},
  \bibinfo{author}{\bibfnamefont{S.~L.} \bibnamefont{{Shapiro}}},
  \bibnamefont{and}
  \bibinfo{author}{\bibfnamefont{K.}~\bibnamefont{{Taniguchi}}},
  \bibinfo{journal}{\prd}  (\bibinfo{year}{2007}), \bibinfo{note}{submitted},
  \eprint{0708.2436}.


\bibitem[{Lor()}]{Lorene}
\bibinfo{note}{{\tt http://www.lorene.obspm.fr/}}.

\bibitem[{\citenamefont{{Tichy} and {Br{\"u}gmann}}(2004)}]{TB}
\bibinfo{author}{\bibfnamefont{W.}~\bibnamefont{{Tichy}}} \bibnamefont{and}
  \bibinfo{author}{\bibfnamefont{B.}~\bibnamefont{{Br{\"u}gmann}}},
  \bibinfo{journal}{\prd} \textbf{\bibinfo{volume}{69}},
  \bibinfo{pages}{024006} (\bibinfo{year}{2004}).

\bibitem[{bho()}]{bhorg}
\bibinfo{note}{{\tt http://www.black-holes.org/researchers3.html}}.

\bibitem[{Cac()}]{Cactus}
\bibinfo{note}{{\tt http://www.cactuscode.org/}}.

\bibitem[{Psi()}]{PsiK}
\bibinfo{note}{We define the normalization of the tetrad to agree with that
  used in \protect\cite{GodRITFP}.}

\bibitem[{\citenamefont{{Baker} et~al.}(2002)\citenamefont{{Baker},
  {Campanelli}, {Lousto}, and {Takahashi}}}]{BCLT}
\bibinfo{author}{\bibfnamefont{J.}~\bibnamefont{{Baker}}},
  \bibinfo{author}{\bibfnamefont{M.}~\bibnamefont{{Campanelli}}},
  \bibinfo{author}{\bibfnamefont{C.~O.} \bibnamefont{{Lousto}}},
  \bibnamefont{and}
  \bibinfo{author}{\bibfnamefont{R.}~\bibnamefont{{Takahashi}}},
  \bibinfo{journal}{\prd} \textbf{\bibinfo{volume}{65}},
  \bibinfo{pages}{124012} (\bibinfo{year}{2002}).

\bibitem[{\citenamefont{{Duez} et~al.}(2003)\citenamefont{{Duez}, {Marronetti},
  {Shapiro}, and {Baumgarte}}}]{DMSB}
\bibinfo{author}{\bibfnamefont{M.~D.} \bibnamefont{{Duez}}},
  \bibinfo{author}{\bibfnamefont{P.}~\bibnamefont{{Marronetti}}},
  \bibinfo{author}{\bibfnamefont{S.~L.} \bibnamefont{{Shapiro}}},
  \bibnamefont{and} \bibinfo{author}{\bibfnamefont{T.~W.}
  \bibnamefont{{Baumgarte}}}, \bibinfo{journal}{\prd}
  \textbf{\bibinfo{volume}{67}}, \bibinfo{pages}{024004}
  (\bibinfo{year}{2003}).

\bibitem[{\citenamefont{{Brown} et~al.}(2007)\citenamefont{{Brown}, {Sarbach},
  {Schnetter}, {Tiglio}, {Diener}, {Hawke}, and {Pollney}}}]{turducken}
\bibinfo{author}{\bibfnamefont{D.}~\bibnamefont{{Brown}}},
  \bibinfo{author}{\bibfnamefont{O.}~\bibnamefont{{Sarbach}}},
  \bibinfo{author}{\bibfnamefont{E.}~\bibnamefont{{Schnetter}}},
  \bibinfo{author}{\bibfnamefont{M.}~\bibnamefont{{Tiglio}}},
  \bibinfo{author}{\bibfnamefont{P.}~\bibnamefont{{Diener}}},
  \bibinfo{author}{\bibfnamefont{I.}~\bibnamefont{{Hawke}}}, \bibnamefont{and}
  \bibinfo{author}{\bibfnamefont{D.}~\bibnamefont{{Pollney}}},
  \bibinfo{journal}{ArXiv e-prints}  (\bibinfo{year}{2007}),
  \eprint{0707.3101}.

\bibitem[{\citenamefont{{Taniguchi} et~al.}(2006)\citenamefont{{Taniguchi},
  {Baumgarte}, {Faber}, and {Shapiro}}}]{TBFS1}
\bibinfo{author}{\bibfnamefont{K.}~\bibnamefont{{Taniguchi}}},
  \bibinfo{author}{\bibfnamefont{T.~W.} \bibnamefont{{Baumgarte}}},
  \bibinfo{author}{\bibfnamefont{J.~A.} \bibnamefont{{Faber}}},
  \bibnamefont{and} \bibinfo{author}{\bibfnamefont{S.~L.}
  \bibnamefont{{Shapiro}}}, \bibinfo{journal}{\prd}
  \textbf{\bibinfo{volume}{74}}, \bibinfo{pages}{041502(R)}
  (\bibinfo{year}{2006}).

\bibitem[{\citenamefont{{Taniguchi} et~al.}(2007)\citenamefont{{Taniguchi},
  {Baumgarte}, {Faber}, and {Shapiro}}}]{TBFS2}
\bibinfo{author}{\bibfnamefont{K.}~\bibnamefont{{Taniguchi}}},
  \bibinfo{author}{\bibfnamefont{T.~W.} \bibnamefont{{Baumgarte}}},
  \bibinfo{author}{\bibfnamefont{J.~A.} \bibnamefont{{Faber}}},
  \bibnamefont{and} \bibinfo{author}{\bibfnamefont{S.~L.}
  \bibnamefont{{Shapiro}}}, \bibinfo{journal}{\prd}
  \textbf{\bibinfo{volume}{75}}, \bibinfo{pages}{084005}
  (\bibinfo{year}{2007}).

\bibitem[{\citenamefont{{Grandcl{\'e}ment}}(2006)}]{Grand}
\bibinfo{author}{\bibfnamefont{P.}~\bibnamefont{{Grandcl{\'e}ment}}},
  \bibinfo{journal}{\prd} \textbf{\bibinfo{volume}{74}},
  \bibinfo{pages}{124002} (\bibinfo{year}{2006}).

\bibitem[{\citenamefont{{Shibata} and {Uryu}}(2007)}]{STU1}
\bibinfo{author}{\bibfnamefont{M.}~\bibnamefont{{Shibata}}} \bibnamefont{and}
  \bibinfo{author}{\bibfnamefont{K.}~\bibnamefont{{Uryu}}},
  \bibinfo{journal}{Class.~Quant.~Grav.} \textbf{\bibinfo{volume}{24}},
  \bibinfo{pages}{125} (\bibinfo{year}{2007}).

\bibitem[{\citenamefont{{Shibata} and {Uryu}}(2006)}]{STU2}
\bibinfo{author}{\bibfnamefont{M.}~\bibnamefont{{Shibata}}} \bibnamefont{and}
  \bibinfo{author}{\bibfnamefont{K.}~\bibnamefont{{Uryu}}},
  \bibinfo{journal}{\prd} \textbf{\bibinfo{volume}{74}},
  \bibinfo{pages}{121503} (\bibinfo{year}{2006}).

\end{thebibliography}
\end{document}